# HALF-METALLIC FERROMAGNETS AND SPIN GAPLESS SEMICONDUCTORS

V. V. Marchenkov*, **, N. I. Kourov*, V. Yu. Irkhin*

*M.N. Mikheev Institute of Metal Physics, UB RAS, 620108 Ekaterinburg, Russia
**Ural Federal University, 620002 Ekaterinburg, Russia
e-mail: march@imp.uran.ru

A brief review of experimental and theoretical studies of half-metallic ferromagnets (HMF) and spin gapless semiconductors (SGS) is presented. An important role of non-quasiparticle states owing to electron-magnon scattering in transport properties is discussed. The problem of low-temperature resistivity in HMF is treated in terms of one-magnon and two-magnon scattering processes.

## INTRODUCTION

One of directions in modern physics of magnetic phenomena and condensed state is to find and detect physical properties of new materials for spintronics (see, e.g., [1, 2]). Such materials include half-metallic ferromagnets (HMF) [3] and spin gapless semiconductors (SGS) [4, 5], in which a high degree of spin polarization of the charge carriers can be realized. To date, we know a lot of Heusler alloys, which have HMF and SGS electronic structure, as well as good thermoelectric properties (see, e.g., [1]). Therefore, intensive researches of Heusler alloys have been performed for the last few years, because they can be used in thermoelectric and spintronics devices.

Disadvantages of conventional HMF-based Heusler alloys as the materials for spintronic and thermoelectric devices should be also noted: they possess metallic conductivity and small values of thermoelectric power. Therefore, it is important to study similar systems having properties close to the classical semiconductors. Recently, half-metallic ferromagnetism in high quality single crystals of doped $HgCr_2Se_4$ has been reported [6].

According to the prediction by X. L. Wang [4], SGS have a number of unique properties associated with their unusual band structure. Such materials make it possible to combine the properties of HMF with semiconductor characteristics with the possibility of fine tuning of the magnitude of the energy gap, and, consequently, of controlling electronic properties. The SGS state was observed in Heusler alloys CoFeMnSi, CoFeCrAl, CoMnCrSi, CoFeVSi and FeMnCrSb [7], $Mn_2CoAl$ [8] and $Ti_2MnAl$ [9].

One of the most important characteristics of spintronic materials is the degree of polarization of the current carriers – ideally it should be 100% at room temperature. It is believed that in SGS materials such polarization can be achieved.

Besides, it is interesting to trace the transition from the ordinary (magnetic and nonmagnetic) metallic and semiconductor states to the HMF state, then to the SGS state, and back again.

## HALF-METALLIC FERROMAGNETS

The main feature of HMF is the presence of a gap at the Fermi level for minority (spin-down) electron states and its absence for majority (spin-up) current carriers with a metallic conductivity (Fig. 1a). This feature is usually revealed as a result of band calculations and in experiments measuring optical properties.

Apparently, the HMF energy gap should be manifested in electronic properties, in particular, in electronic transport. According to the mean-field picture, HMF can be regarded as a system of two parallel-connected conductors [10, 11]. One of them is a subsystem of current carriers with spins up, and the other is a subsystem with spins down. The first one, the spin-up subsystem, has a typical "metallic" conductivity, so that its resistivity increases according to a power law with temperature. The spin-down subsystem has a "semiconductor" conductivity so that its resistivity increases according to a power law with temperature. Under certain conditions, its conductivity

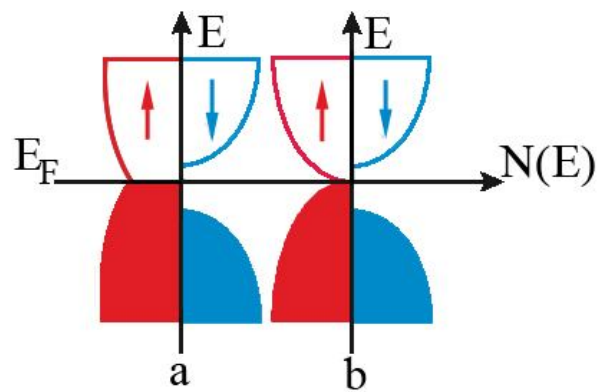

Fig.1. Schematic view of density of states $N(E)$ as a function of energy $E$: (a) half-metallic ferromagnet and (b) spin gapless semiconductor. The occupied states are indicated by filled areas. Arrows indicate the majority (↑) and minority (↓) states.

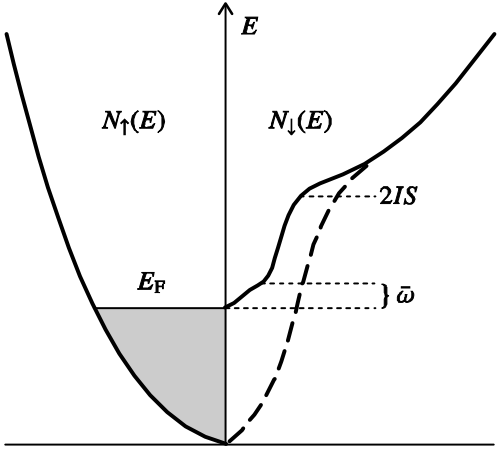

Fig.2. The HMF density of states, 2IS being the spin splitting, $\varpi$ the characteristic magnon frequency. The dashed line shows the finite temperature tail of NQP states which is proportional to $T^{3/2}$.

can have an exponential temperature dependence.

The mean-field picture is considerably modified when including non-quasiparticle (NQP) states owing to electron-magnon scattering [12]. These states violate 100% spin polarization of current carriers: they yield the density of incoherent states in the energy gap, proportional to $T^{3/2}$ (Fig. 2).

The NQP states make considerable contributions to magnetic and transport properties [12]. Spin transport in HMF at finite temperatures was theoretically investigated with account of minority-spin band NQP states in Ref. [13]. Spin Hall conductivity proportional to $T^{3/2}$ was found.

The magnetic resistivity of usual ferromagnetic metals is determined by one-magnon processes which yield

$$\rho(T) \propto T^2 N_\uparrow(E_F) N_\downarrow(E_F) \exp(-T/T^*),$$

where $T^* \propto q_1^2 T_C$ is the characteristic scale for these processes, $q_1 \propto \Delta/v_F$, $\Delta = 2IS$ being the spin splitting (the energy gap). This contribution is absent for HMF since $N_\downarrow(E_F) = 0$.

Two-magnon scattering processes [14, 15] lead to a power-law temperature dependence of the resistivity $\rho(T) \propto T^n$, as well as to a negative linear magnetoresistance. We have $n = 9/2$ at $T<T^{**}$ and $n = 7/2$ at $T>T^{**}$, $T^{**} \propto q_2^2 T_C$. In the simple one-band model of HMF where $E_F < \Delta$ one has $q_2 \propto (\Delta/W)^{1/2}$ with $W$ the bandwidth. Generally speaking, $q_2$ may be sufficiently small provided that the energy gap is much smaller than $W$, which is typical of real HMF systems.

In Ref. [16], the temperature dependences of the resistivity in $Co_2FeSi$ single crystals were studied and the exponential temperature contribution to the resistivity was fitted. The authors of [16] analyze the electrical resistance rather than the conductivity, as it should be in the case for two parallel conduction channels. The gap determined from the experiment is only 103 K (8.9 meV), which is about an order of magnitude lower than a typical value for HMF. The gap determined from the experiment is only 103 K (8.9 meV), which is about an order of magnitude lower than a typical value for HMF. At temperatures on the order of the gap, the contribution from the exponential (gap) term determined by the authors is only a couple of percent as compared to the total resistivity.

The band structure calculations for this compound [17] show that the HFM picture is highly sensitive to lattice parameters, so that the gap can vanish. One can suppose that $Co_2FeSi$ is on the boundary of the HMF state, so that the temperature $T^*$ is high. Most important, the exponential dependence in this case should be determined not by the gap $\Delta$, but by much smaller quantity $T^*$.

To clarify the situation, precision investigations of the field and temperature dependences of the resistivity were carried out in [18] for $Co_2MeZ$ (Me = Ti, V, Cr, Mn, Fe, Ni, and Z = Al, Si, Ga, Ge, In, Sn, Sb).

Fig. 3a shows the temperature dependence of the resistivity of the HMF $Co_2FeSi$ without magnetic field and in a field of 100 kOe. The resistivity $\rho$ can be presented as follows (Fig. 3b)

$$\rho(T) = \rho_0 + AT^2 + BT^n + CT^2,$$

where $\rho_0$ is the residual resistivity, $A$, $B$ and $C$ are the coefficients, and $n$ is a power index. There are three temperature intervals in which the resistivity depends on temperature and magnetic field in different ways (Fig.3b)

1) below 30 K, $\rho(T) \propto T^n$ with $n = 2$, and coefficient $A \propto H^2$; 2) from 30 to 60 K, $\rho(T) \propto T^n$ with $n \approx 4$, and coefficient $B \propto -H$; 3) above 65 K $\rho(T) \propto T^n$ with $n \approx 2$, and coefficient $C \propto H^{-2}$. The experimental results obtained indicate that in the temperature range 30 K < $T$ < 60 K a power-law temperature dependence of the electrical resistivity with a power exponent and a linear negative magnetoresistance is observed. These facts appear to be a manifestation of two-magnon scattering processes: these are the main mechanism for the scattering of current carriers determining the behavior of the electrical and magnetoresistance of the alloy at temperatures of 30 K < T < 60 K.

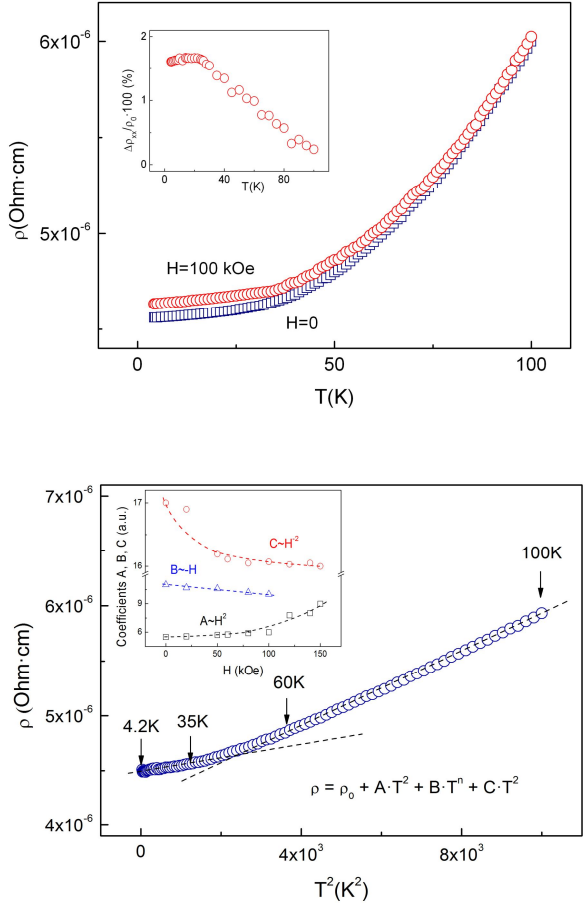

Fig.3. (a) Temperature dependence of resistivity for $Co_2FeSi$ without and in a field of 100 kOe. The insert shows the temperature dependence of the magnetoresistivity. (b) The dependence of resistivity $\rho = f(T^2)$. The field dependences of the coefficients $A$, $B$ and $C$ are shown in insert.

## SPIN GAPLESS SEMICONDUCTORS

SGS have a number of unique properties associated with their unusual band structure, i.e. the presence of a wide ($\Delta E \sim 1$ eV) gap at the Fermi level for carriers with spin down and zero energy gap for spin-up carriers (Fig. 1b). In such SGS materials, strong ferromagnetism is expected with a high Curie temperature and a 100% spin polarization of the charge carriers at room temperature. Consequently, they can be convenient objects for practical use in thermoelectric devices and spintronics devices. Band calculations [8, 19-21] and experimental studies of transport [8, 22], magnetic [8, 19, 21, 22], and optical [23] properties have shown that this class of materials includes the alloy $Mn_2CoA$ ordered in a structure of the type $F\bar{4}3m$ ($Hg_2CuTi$). According to [21], at temperatures below $T_C = 720$ K this alloy has ferromagnetic ordering, and its ground-state magnetic moment is equal to 2 $\mu_B$/f.u.

It was shown in [8, 19] that when the ratio of the lattice constants $c/a$ in $Mn_2CoAl$ decreases, the energy gap in the electron spectrum is completely closed. Moreover, even for a $c/a$ ratio corresponding to normal conditions, any type of disorder (mutual substitution of Mn, Co or Al atoms) leads to an increase in the density of states for spin-up electrons, whereas for spin-down electrons the energy gap is substantially reduced or in some cases even closed. To verify the results of [8, 19], optical properties were measured and the electronic band structure of $Mn_2CoAl$-based alloys was calculated in [23], both near the stoichiometric composition and with deviation from stoichiometry. An anomalous behavior of the optical properties of the alloy $Mn_{1.8}Co_{1.2}Al$ was observed, i.e. positive values of the dielectric-constant real part $\varepsilon_1$ and the absence of a Drude contribution to the optical conductivity in the infrared region of the spectrum [23]. This indicates a weakening of the metallic properties of the alloy. As a result of the calculations, a picture of the band spectrum was presented, which is characteristic of spin gapless semiconductors and makes it possible to explain the features of the optical absorption spectrum.

## ACKNOWLEDGMENTS

The work was carried out within the framework of the state assignment of FASO of Russia (the themes "Spin" No. AAAA-A18-118020290104-2 and "Quant" No. AAAA-A18-118020190095-4) with partial support of the RFBR (project No. 18-02-00739) and the Government of the Russian Federation Federation (ordinance No. 211, contract No. 02.A03.21.0006).